\documentclass[aps,prl,twocolumn,superscriptaddress]{revtex4-2}
\usepackage{ulem,graphicx,amssymb,amsfonts,amsmath,chemarr,color,commath,braket,hyperref}

\newcommand{\beginsupplement}{%
\setcounter{page}{1}
        \setcounter{equation}{0}
        \renewcommand{\theequation}{S\arabic{equation}}%
        \setcounter{table}{0}
        \renewcommand{\thetable}{S\arabic{table}}%
        \setcounter{figure}{0}
        \renewcommand{\thefigure}{S\arabic{figure}}%
     }

\begin{document}

\title{Effects of molecular noise on cell size control}

\author{Motasem ElGamel}
\affiliation{Department of Physics and Astronomy, University of Pittsburgh, Pittsburgh, Pennsylvania 15260, USA}

\author{Andrew Mugler}
\email{andrew.mugler@pitt.edu}
\affiliation{Department of Physics and Astronomy, University of Pittsburgh, Pittsburgh, Pennsylvania 15260, USA}

\begin{abstract}
Cells employ control strategies to maintain a stable size. Dividing at a target size (the `sizer' strategy) is thought to produce the tightest size distribution. However, this result follows from phenomenological models that ignore the molecular mechanisms required to implement the strategy. Here we investigate a simple mechanistic model for exponentially growing cells whose division is triggered at a molecular abundance threshold. We find that size noise inherits the molecular noise and is consequently minimized not by the sizer but by the `adder' strategy, where a cell divides after adding a target amount to its birth size. We derive a lower bound on size noise that agrees with publicly available data from six microfluidic studies on {\it Escherichia coli} bacteria. 
\end{abstract}

\maketitle

Maintaining a stable cell size is a central requirement of life. Fatal consequences to large cell size fluctuations include cytoplasm dilution \cite{neurohr2019excessive} and impaired mitochondrial function \cite{miettinen2016cellular}. Additionally, cell size is important for optimizing nutrient intake \cite{chien2012cell, turner2012cell}, accommodating intracellular content \cite{turner2012cell, marshall2012determines}, maintaining uniformity in tissues \cite{ginzberg2015being}, and more \cite{young2006selective}. Size stability, in exponentially growing cells, does not emerge passively: because of unavoidable noise in growth and division, cells employ active size control strategies \cite{chien2012cell, turner2012cell, ginzberg2015being}. The strategy predicted to produce the tightest cell size distribution is known as the `sizer' \cite{amir2014cell, facchetti2017controlling}. In this strategy, a cell divides when a target size is reached, regardless of its birth size or the required growth time. Because the sizer attempts to reset the cell size every generation, it makes sense that this strategy would lead to minimal size noise. Yet, pure sizers are rarely observed in microbial growth control.

The prediction that a sizer has the lowest size noise is based on phenomenological models that ignore underlying molecular mechanisms \cite{amir2014cell, tanouchi2015noisy, willis2017sizing, susman2018individuality}. Dividing at a target size requires a molecular mechanism that tells the cell when the target is reached, and that mechanism may have its own noise that impacts size noise. Indeed, molecular noise has been shown to have important effects on cell size control, even in a high gene expression regime \cite{sassi2022protein, biswas2022cell}. Noise in the accumulation of a division-triggering molecule can explain the universality of size distributions in the `adder' strategy \cite{ghusinga2016mechanistic}, where a cell divides after adding a target amount to its birth size \cite{sompayrac1973autorepressor, voorn1993mathematics, amir2014cell, sauls2016adder}. Noise in the accumulation threshold itself contributes to size noise and can even alter the observed strategy among sizer, adder, and `timer' (where a cell divides after a target time) \cite{luo2023stochastic}. Molecular noise in the DNA replication mechanism \cite{berger2022robust} or cell-to-cell variability \cite{facchetti2019reassessment} can make sizer control appear adder-like. Together, these works show that molecular noise has a driving impact on cell size control, but a simple and mechanistic understanding of its effects on cell size noise across the timer-adder-sizer spectrum remains elusive.

Here we introduce a mechanistic model of cell size control in which division occurs when a single molecular species (such as FtsZ \cite{chien2012cell, si2019mechanistic} or peptidoglycan \cite{harris2016relative}) accumulates to an abundance threshold. The model admits the timer, adder, and sizer as limits, and we find that the variance in birth size is minimized by the adder, not the sizer. The reason is that the sizer mechanism requires active protein degradation in our model, resulting in high molecular noise for a fixed protein production cost, and this noise overpowers the sizer's otherwise tight control. We predict a lower bound on size noise that is lowest for the adder and find agreement with publicly available data from six microfluidic studies on {\it Escherichia coli} bacteria.

\begin{figure}[b]
\includegraphics[width=.9\columnwidth]{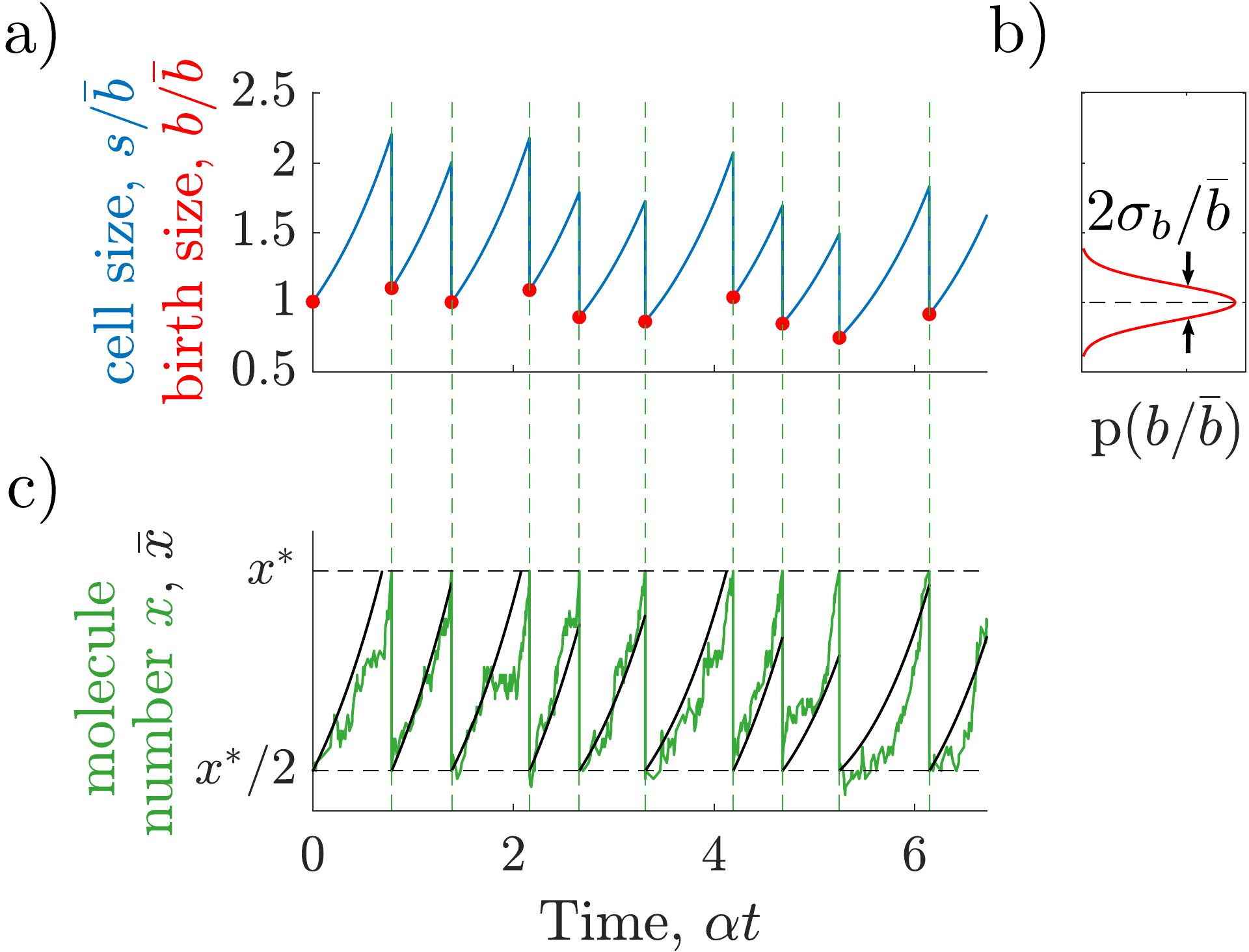}
\caption{(a) A cell grows exponentially and divides in half. (b) The birth size fluctuates. (c) Division occurs when a molecule reaches an abundance threshold. Noise in the molecule number contributes to noise in the birth size. Here $\gamma=10^{-2}$, $\rho = 1$, and $k/\alpha = 50$.}
\label{model}
\end{figure}

We first summarize the prevailing phenomenological model of cell size control \cite{amir2014cell, susman2018individuality}. The simplest form assumes that a cell grows exponentially at a constant rate and divides in half [Fig.\ \ref{model}(a)]. In the $n$th generation, a cell with birth size $b_n$ and growth rate $\alpha$ has size
\begin{equation}
\label{s}
s_n(t) = b_ne^{\alpha t}
\end{equation}
at time $t$. Denoting the division time as $t_n$, the new birth size is $b_{n+1} = b_ne^{\alpha t_n}/2$. Defining $\epsilon_n=\ln(b_n/\bar{b})$ as the logarithmic deviation of the birth size from its long-time average, and $\delta_n=\alpha t_n-\ln 2$ as the deviation of the exponential phase from its expected value for size doubling, this expression becomes
\begin{equation}
\label{epsilon}
    \epsilon_{n+1}=\epsilon_n+\delta_n.
\end{equation}
If $\delta_n$ is independent of $\epsilon_n$, then Eq.\ \ref{epsilon} describes a random walk, which is not stable. Therefore, most size control models assume that the phase corrects for deviations in the birth size \cite{osella2014concerted, tanouchi2015noisy, susman2018individuality},
\begin{equation}
\label{delta}
    \delta_n=-\beta\epsilon_n+\eta_n.
\end{equation}
Here, the homeostasis parameter $\beta$ sets the strength of the correction, and $\eta_n$ is uncorrelated Gaussian noise. Eq.\ \ref{delta} ensures that cells born larger ($\epsilon_n > 0$) grow for less time ($\delta_n < 0$) on average.

The values $\beta = 0$, $1/2$, and $1$ correspond to the timer, adder, and sizer strategies, respectively \cite{amir2014cell, susman2018individuality}. Correspondingly, $\beta$ controls the noise in the birth size, $\sigma^2_b/\bar{b}^2$ [Fig.\ \ref{model}(b)]. Specifically, experiments in bacteria suggest $\sigma_b/\bar{b} \sim 20\%$ \cite{wang2010robust, campos2014constant, taheri2015cell, wallden2016synchronization, si2019mechanistic, vashistha2021non}, for which $\sigma^2_\epsilon \approx \sigma^2_b/\bar{b}^2 \ll 1$. Inserting Eq.\ \ref{epsilon} into Eq.\ \ref{delta} and considering the variance obtains $\sigma^2_\epsilon = (1-\beta)^2\sigma^2_\epsilon + \sigma^2_\eta$ in steady state. Solving for $\sigma^2_\epsilon$, we see that the size noise,
\begin{equation}
\label{size_noise}
\frac{\sigma^2_b}{\bar{b}^2} \approx \sigma^2_\epsilon = \frac{\sigma^2_\eta}{\beta(2-\beta)},
\end{equation}
is minimized for the sizer at $\beta = 1$.

In Eq.\ \ref{delta}, the homeostasis parameter $\beta$ and the timing noise $\eta_n$ are phenomenological, rather than arising from an underlying molecular mechanism. Our key advance will be to show that the mechanism that sets $\beta$ also affects $\eta_n$, such that the two are not independent as commonly assumed. Instead, we will see that the coupling between $\beta$ and $\eta_n$ endows $\sigma^2_\eta$ in Eq.\ \ref{size_noise} with an effective $\beta$ dependence, opening the possibility that the sizer does not minimize size noise after all.

Consider a molecular species whose abundance $x$ triggers cell division when it reaches a threshold $x_*$ [Fig.\ \ref{model}(c)]. We intend this construction to be minimal and generic \cite{teather1974quantal, ghusinga2016mechanistic}, but we are also motivated by specific molecular species in bacteria such as FtsZ \cite{chien2012cell, si2019mechanistic} or peptidoglycan \cite{harris2016relative} that are thought to accumulate to a threshold amount to initiate division. We assume that the threshold is fixed and focus on the timing noise in reaching it, rather than preexisting noise in its value \cite{luo2023stochastic}.
For simplicity we ignore the initiation of DNA replication, which is also thought to be an important trigger for cell division and can affect size control \cite{berger2022robust}.

We prescribe the simplest possible reactions for $x$, namely linear production and degradation. We will see that allowing production to either scale with \cite{harris2016relative, si2019mechanistic} or be independent of cell size will allow the model to reduce to the timer, adder, and sizer strategies in particular limits. Thus, the dynamics of $x$ within generation $n$ are
\begin{equation}
\label{xdot}
\frac{d\bar{x}_n}{dt} = \nu + \mu s_n - \lambda\bar{x}_n,
\end{equation}
where $\nu$ is the size-independent production rate, $\mu s_n$ is the size-dependent production rate, $\lambda$ is the degradation rate, and the bar denotes the fact that we will later be interested in the noise in $x$. Although Eq.\ \ref{xdot} is not the only model that spans the timer-adder-sizer spectrum \cite{nieto2020unification},
we are motivated by experiments that specifically suggest that degradation \cite{si2019mechanistic} and size-proportional production \cite{harris2016relative, si2019mechanistic} are responsible for sizer and adder control, respectively, as we will see for our model below. For simplicity and consistency with the phenomenological model above, we neglect the effects of nonexponential growth \cite{kar2021distinguishing, cylke2022super}, heterogeneous growth rates \cite{kohram2021bacterial, biswas2022cell}, and noisy \cite{susman2018individuality} or asymmetric division \cite{campos2014constant, iyer2014scaling, barber2021modeling} (although we relax the latter two assumptions later on). We further assume that $x$ is initialized at $x_*/2$ each generation, corresponding to symmetric partitioning at division, although none of our conclusions change if instead $x$ is initialized at zero, for example if the molecule is cleared or used in pole construction \cite{harris2016relative}.

\begin{figure}
\includegraphics[width=\columnwidth]{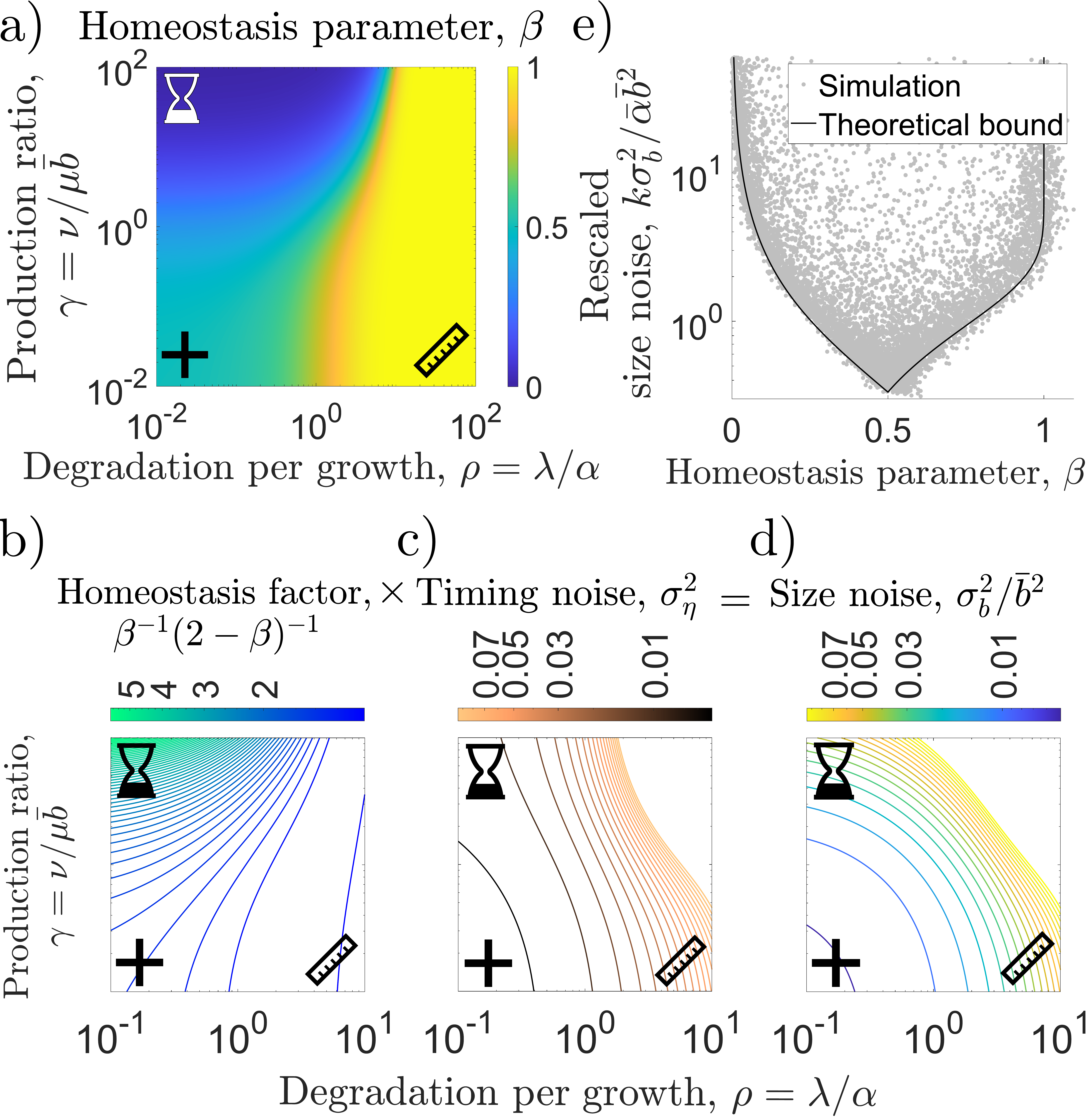}
\caption{(a) Homeostasis parameter $\beta$ is a function of mechanistic parameters $\gamma$ and $\rho$ in our model. Symbols indicate limiting cases of timer (upper left), adder (lower left), and sizer (lower right). (b-d) Dependence of each component of the size noise on $\gamma$ and $\rho$. (e) Rescaled size noise ($CV^2$) vs.\ homeostasis parameter $\beta$ from simulations.
}
\label{mean}
\end{figure}

If $\mu = \lambda = 0$ in Eq.\ \ref{xdot}, then $\bar{x}_n(t) = x_*/2 + \nu t$, which reaches $x_*$ in a constant time, corresponding to the timer strategy. If instead $\nu = \lambda = 0$, then $\bar{x}_n(t) = x_*/2 + \mu b_n(e^{\alpha t}-1)/\alpha$ using Eq.\ \ref{s}. Solving the division condition $\bar{x}_n(t) = x_*$ for $t$ and inserting it into Eq.\ \ref{s} obtains $s_n = b_n + \alpha x_*/2\mu$, which shows that the cell adds a constant amount to its birth size---the adder strategy \cite{harris2016relative, si2019mechanistic}. Finally, if only $\nu=0$, then Eq.\ \ref{xdot} reads $d\bar{x}_n/dt = \mu s_n - \lambda\bar{x}_n$. If degradation is much faster than cell growth, $\lambda\gg\alpha$, then $s_n(t)$ is quasi-static on the response timescale of $x$, and $\bar{x}_n(t) \approx \mu s_n(t)/\lambda$. Thus, a molecule number threshold is equivalent to a size threshold, corresponding to the sizer strategy. These three limits suggest that we define two dimensionless parameters, $\gamma = \nu/\mu\bar{b}$ and $\rho = \lambda/\alpha$, for which the timer, adder, and sizer correspond to
$\{\gamma\gg1, \rho\ll1\}$, $\{\gamma\ll1, \rho\ll1\}$, and $\{\gamma\ll1, \rho\gg1\}$,
respectively, as illustrated by the icons in the corners of Fig.\ \ref{mean}(a). For reference, a complete list of parameter definitions is given in \cite{supp}.

In our model, the homeostasis parameter $\beta$ defined by Eq.\ \ref{delta} is a function of the mechanistic parameters $\gamma$ and $\rho$. To see this, we write the general solution to Eq.\ \ref{xdot},
$\bar{x}_n(t) = x_*e^{-\rho\alpha t}/2 + (k/\alpha)[(b_n/\bar{b})(e^{\alpha t}-e^{-\rho\alpha t})/(1+\rho) + \gamma(1-e^{-\rho\alpha t})/\rho]/(1+\gamma)$.
Here we have defined $k=\nu+\mu\bar{b}$ as the total molecule production rate. It represents the intrinsic biochemical rate at which a molecule is produced, and therefore we keep it fixed throughout. Fixing $k$ is consistent with observed dependences of constitutive gene expression \cite{klumpp2014bacterial} (in the timer limit) and of balanced biosynthesis \cite{harris2016relative, si2019mechanistic} (in the adder limit) on the cell growth rate. Nevertheless, we find that our conclusions are unchanged if we instead fix the threshold $x_*$ \cite{supp}.

To find $\beta$ from $\bar{x}_n(t)$, we again take $\epsilon_n=\ln(b_n/\bar{b})$ to be small, and we consider times $t$ near division, where $\delta=\alpha t-\ln 2$ is expected to be small. We expand the expression for $\bar{x}_n(t)$ to linear order in $\epsilon_n$ and $\delta$ as
\begin{equation}
\label{expand}
\bar{x}_n(t) \approx c_0 + c_1\epsilon_n + c_2\delta,
\end{equation}
where the expansion coefficients $c_0$, $c_1$, and $c_2$ are functions of $x_*$, $\gamma$, $\rho$, and $k/\alpha$ \cite{supp}.
At division, we have $\bar{x}_n(t) = x_*$ and $\delta=\delta_n$. The constant terms in Eq.\ \ref{expand} then read $x_* = c_0$, which when solved for $x_*$ obtains
\begin{equation}
\label{xstar}
x_* = 2\left(\frac{k/\alpha}{1+\gamma}\right)\left[\frac{1}{1+\rho}
	+ \frac{\gamma}{\rho}\left(\frac{r-1}{2r-1}\right)\right],
\end{equation}
where $r\equiv2^\rho$. Eq.\ \ref{expand} then reads $\delta_n = -(c_1/c_2)\epsilon_n$, which when compared with Eq.\ \ref{delta} implies
\begin{equation}
\label{beta}
\beta = \frac{c_1}{c_2} = \frac{(2r-1)^2}{4r^2+(g-2)r},
\end{equation}
where $g\equiv\gamma\rho+\gamma$, and the second step includes inserting Eq.\ \ref{xstar} into the expression for $c_2$. Eq.\ \ref{beta} is plotted in Fig.\ \ref{mean}(a), and we see that, as expected, $\beta$ approaches $0$, $1/2$, and $1$ in the timer, adder, and sizer limits, respectively.

In principle, having calculated $\beta$ for our model, Eq.\ \ref{size_noise} would then give the size noise. The factor $\beta^{-1}(2-\beta)^{-1}$ from Eq.\ \ref{size_noise}, which we call the homeostasis factor, is plotted in Fig.\ \ref{mean}(b), and we see that it is smallest for the sizer and largest for the timer, as commonly expected. However,  thus far we have ignored noise in $x$. Noise in $x$ will propagate to noise in division timing and, in turn, to noise in cell size \cite{ghusinga2016mechanistic} [Fig.\ \ref{model}(c)]. To see this, we calculate in our model the statistics of the noise term $\eta_n$ defined by Eq.\ \ref{delta}. Specifically, the timing noise is
\begin{equation}
\label{timing_noise}
\sigma^2_{\eta} = \left\langle\sigma^2_{\delta_n|\epsilon_n}\right\rangle
\approx \left\langle\left(\left.\frac{\partial \bar{x}_n}{\partial\delta}\right|_{\delta=0}\right)^{-2}\sigma^2_{x_n|\epsilon_n}\right\rangle
= \frac{\left\langle\sigma^2_{x_n|\epsilon_n}\right\rangle}{c_2^2}.
\end{equation}
The first step follows from Eq.\ \ref{delta}, conditioned on birth size, where the average is over birth size. The second step approximates the division noise (the noise in the first-passage time for $x_n$ to reach $x_*$) by the molecule number noise, propagated via derivative. The third step takes this derivative from Eq.\ \ref{expand}. We solve for the molecule number noise from the master equation \cite{supp} and find that it varies between the Poissonian limits of $\langle\sigma^2_{x_n|\epsilon_n}\rangle = x_*/2$ for $\rho\ll1$ and $\langle\sigma^2_{x_n|\epsilon_n}\rangle = x_*$ for $\rho\gg1$
\cite{Note1}.
Inserting it into Eq.\ \ref{timing_noise} gives the timing noise, plotted in Fig.\ \ref{mean}(c). We see that the timing noise is largest for the sizer. The reason is that the sizer requires strong degradation ($\rho\gg1$), which, at a fixed production rate $k$, corresponds to fewer total molecules. Indeed, Eq.\ \ref{xstar} shows that $x_*\to0$ as $\rho\to\infty$. A lower threshold $x_*$ is reached in fewer sequential steps, corresponding to larger timing noise.

The size noise  is the product of the homeostasis factor and the timing noise (Eq.\ \ref{size_noise}). Using Eqs.\ \ref{beta} and \ref{timing_noise}, $\sigma^2_b/\bar{b}^2 \approx \langle\sigma^2_{x_n|\epsilon_n}\rangle/[c_1(2c_2-c_1)]$. Inserting the molecule number noise and expansion coefficients and simplifying \cite{supp},
\begin{equation}
\label{size_noise2}
\frac{\sigma^2_b}{\bar{b}^2} = \frac{\alpha}{k}\left[\frac{(1+\gamma)(1+\rho)(2r^2-1)[(g+2\rho)r-(g+\rho)]}
		{\rho[8r^3+4(g-1)r^2-2(g+1)r+1]}\right],
\end{equation}
where again $r\equiv2^\rho$ and $g\equiv\gamma\rho+\gamma$. Eq.\ \ref{size_noise2} is plotted in Fig.\ \ref{mean}(d), and we see that it is minimized for the adder. The reason is that the homeostasis factor is largest for the timer [Fig.\ \ref{mean}(b)], whereas the timing noise is largest for the sizer [Fig.\ \ref{mean}(c)], and this tradeoff makes their product smallest in between, for the adder.
We have checked that Eqs.\ \ref{beta} and \ref{size_noise2} agree with growth-and-division simulations, with division driven by  stochastic reactions corresponding to the terms in Eq.\ \ref{xdot} \cite{gillespie1977exact}.

Because Eqs.\ \ref{beta} and \ref{size_noise2} each depend on at least two parameters, there is no unique function relating the observables $\sigma_b^2/\bar{b}^2$ and $\beta$. However, there is a lower bound. The lower bound is obtained by solving Eq.\ \ref{beta} for $\gamma$, inserting the solution into Eq.\ \ref{size_noise2}, and minimizing with respect to $\rho$. We find numerically that the minimum corresponds to $\rho\to0$ when $0<\beta\le1/2$ and to $\gamma\to0$ when $1/2<\beta<1$. In these limits, Eq.\ \ref{size_noise2} becomes
\begin{equation}
\label{min}
\frac{\sigma^2_b}{\bar{b}^2} \ge \frac{\alpha/k}{\beta(2-\beta)}
\begin{cases}
(1-\beta)[\beta+(1-2\beta)\ln2] & \beta\le1/2 \\
c(2\beta^2-4\beta+1)\ln(1-\beta) & \beta>1/2,
\end{cases}
\end{equation}
where $c\equiv(2\ln2)^{-1}$. Eq.\ \ref{min} is smallest for the adder ($\beta = 1/2$), giving $\sigma_b^2/\bar{b}^2\ge\alpha/3k$. Eq.\ \ref{min} also makes clear that size noise decreases for smaller $\alpha$ or larger $k$, either of which allows more molecules to be produced in a generation. Finally, the denominator in Eq.\ \ref{min} is the homeostasis factor $\beta^{-1}(2-\beta)^{-1}$. Comparing with Eq.\ \ref{size_noise}, this fact makes clear that the molecular mechanism has endowed the timing noise $\sigma^2_\eta$ with a $\beta$ dependence, i.e., the numerator in Eq.\ \ref{min}.

We test Eq.\ \ref{min} against our simulations \cite{supp} in Fig.\ \ref{mean}(e). Each point corresponds to a different value of $\gamma$, $\rho$ and $\alpha/k$, sampled uniformly in log space. We see that the simulated data points obey a lower bound on rescaled size noise $k\sigma_b^2/\alpha\bar{b}^2$ at each $\beta$ value, in good agreement with Eq.\ \ref{min}, with minor discrepancy due to the approximations we made in Eqs.\ \ref{expand} and \ref{timing_noise}. We also test the robustness of our results to other typical noise sources, including growth rate variability, molecule partitioning noise, and noise in the molecular abundance threshold $x^*$ \cite{modi2017analysis} (Fig.\ \ref{noise_robust} \cite{supp}). We find that adding noise sources generally increases size noise levels, as expected. Moreover, we find that noise in $x^*$, depending on the correlation time of fluctuations, can shift the data towards the timer (for large correlation time) or the sizer (for small correlation time), consistent with previous results \cite{luo2023stochastic}. In all cases, our predicted bound is obeyed, and a clear minimum in size noise exists away from the sizer.

\begin{figure}
\includegraphics[width=1\columnwidth]{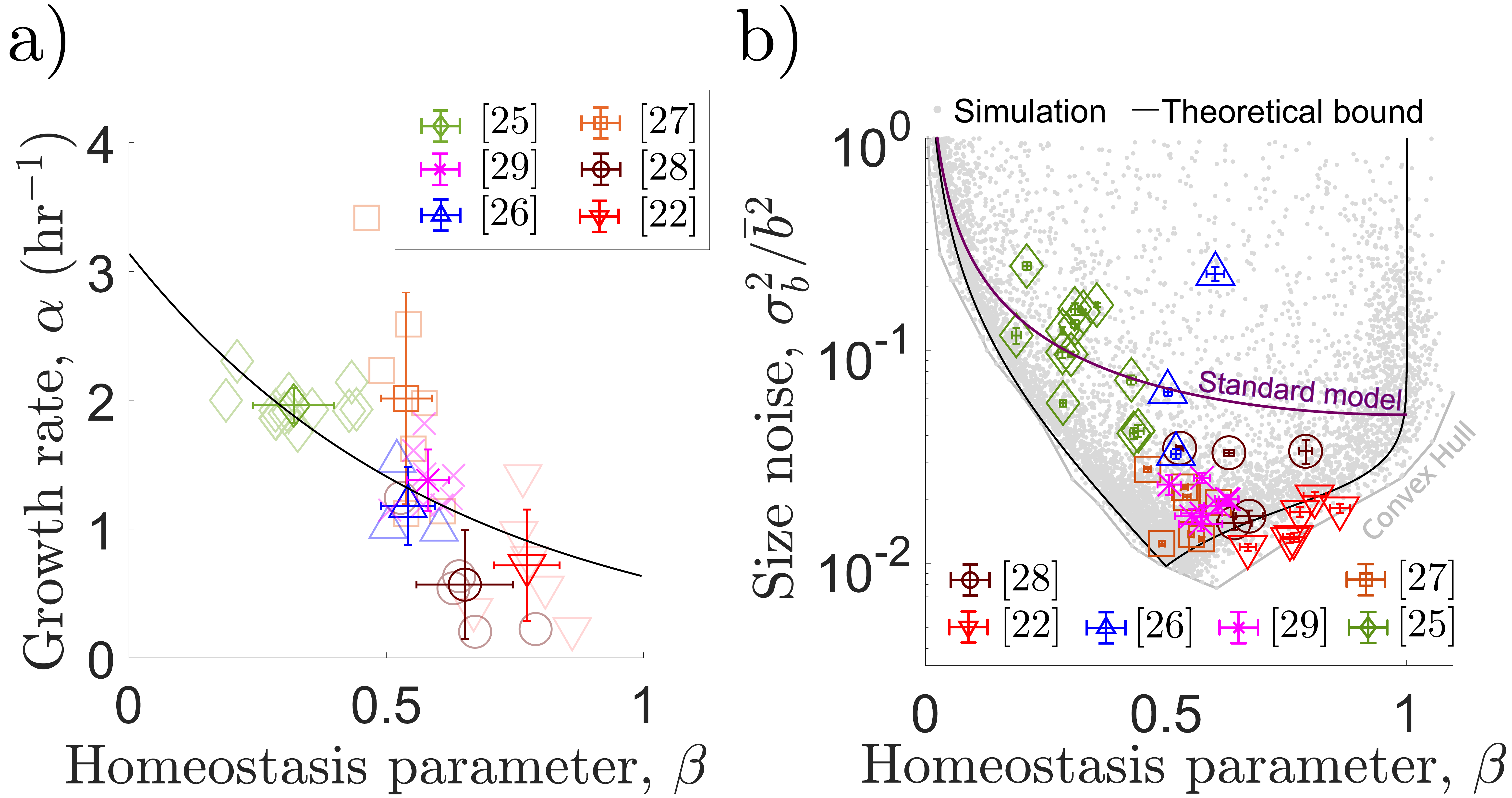}
\caption{(a) Growth rate $\alpha$ vs.\ homeostasis parameter $\beta$ from publicly available data. Data fit to $\alpha(\beta)=3.1 \exp{(-1.6\beta)}$ hr$^{-1}$ (black line). (b) Size noise ($CV^2$) vs.\ $\beta$ from data in a, compared to theoretical lower bound (Eq.\ \ref{min}) and simulations, with $\alpha(\beta)$ inserted, and to the best-fit standard model (Eq.\ \ref{size_noise}). In b, $k = 0.8$/min, set such that convex hull of simulation points first intersects the data.}
\label{noise}
\end{figure}

Since Eq.\ \ref{min} depends on $\alpha$, to compare our theory with experiments, we must take the dependency of $\alpha$ on $\beta$ into account. 
Because our theory does not probe $\alpha$ directly, but rather the ratio $\rho=\lambda/\alpha$,
we rely on experimental data to determine the $\alpha$-$\beta$ relation empirically. Fig.\ \ref{noise}(a) shows publicly available data from six microfluidic studies on \textit{E.\ coli} \cite{wang2010robust, campos2014constant, taheri2015cell, wallden2016synchronization, si2019mechanistic, vashistha2021non} (see \cite{supp} for data analysis). We see that $\beta$ generally decreases with $\alpha$ across studies, a trend that is widely observed \cite{tanouchi2015noisy,si2019mechanistic,wallden2016synchronization}. We fit the data in Fig.\ \ref{noise}(a) to an exponentially decaying function, resulting in $\alpha=3.1 \exp{(-1.6\beta)}$ hr$^{-1}$ (black line).

Inserting this dependence into Eq.\ \ref{min} gives the lower bound shown in Fig.\ \ref{noise}(b) (black line), along with the corresponding simulation data and their convex hull shown in gray. We compare this prediction to experimental size noise data from the same six studies. We see that the theory explains the data, specifically the strong fall-off of the size noise with $\beta$ in the timer-adder region, the minimum near the adder, and the increase of noise with $\beta$ in the adder-sizer region (in particular the data from \cite{si2019mechanistic}, although more data would be needed at large $\beta$ to verify this increase). In contrast, we see that the best fit of the standard model (Eq.\ \ref{size_noise}, purple) fails to explain these features and is a poorer description of the data. Note that to set $k$ in Eq.\ \ref{min}, we decrease it (thus increasing the predicted noise bound) until the simulation convex hull first intersects the data in Fig.\ \ref{noise}(b). The resulting value of $k\approx1$/min is a plausible rate of protein production \cite{kennell1977transcription} and corresponds to a copy number of at least $50$$-$$500$ molecules per cell
\cite{Note2}.
Consistently, experimental estimates of the number of FtsZ proteins per cell are in the thousands \cite{feucht2001cytological}.

We have demonstrated, using a minimal model of threshold-triggered division in bacteria, that cell size noise is minimized by the adder strategy, not the sizer strategy as conventionally expected. The reason is that molecular noise, missing in the conventional framework, amplifies size noise in the sizer limit, defined in our model by active protein degradation as suggested in experiments \cite{si2019mechanistic}. The amplification is due to high timing noise [Fig.\ \ref{mean}(c, d)], consistent with recent related work \cite{proulx2022evolution}. Our predictions are supported by data from six studies in {\it E.\ coli} \cite{wang2010robust, campos2014constant, taheri2015cell, wallden2016synchronization, si2019mechanistic, vashistha2021non} [Fig.\ \ref{noise}(b)]. Specifically, we find that while the data span a range of $\beta$, for a given $\beta$ most data lie close to the predicted noise bound, with exceptions that may be due to variations from other noise sources (Fig.\ \ref{noise_robust} \cite{supp}). This suggests that size noise might not be minimized globally, but rather, for a given size control strategy, minimized for that strategy.
Additionally, we predict that if cells are forced deeply into the sizer regime, either by slowing growth \cite{wallden2016synchronization, si2019mechanistic} or perturbing degradation \cite{si2019mechanistic}, size noise should increase, not decrease as predicted by the standard model [Fig.\ \ref{noise}(b)].

Most bacteria exhibit adder control \cite{campos2014constant, taheri2015cell, sauls2016adder, willis2017sizing}, raising the question of whether the adder is optimal in some sense \cite{lin2017effects, proulx2022evolution}. Our model suggests that the adder, not the sizer, may provide the tightest attainable size control for bacteria.
Other organisms show different size control mechanisms, with fission yeast, for example, exhibiting a strong sizer \cite{sveiczer1996size}. In fission yeast, division timing depends on a concentration threshold rather than a molecule number threshold as studied here \cite{wood2015sizing}. We leave concentration-dependent size control for future work. 

Our work emphasizes that the molecular mechanism underpins not only the size control strategy, but its statistics as well. Although we have focused on size noise in this work, we anticipate that this idea will have consequences for other questions traditionally informed by a phenomenological understanding of size control, including multigenerational memory \cite{elgamel2022multigenerational}, cell geometry \cite{facchetti2019reassessment}, population-level effects \cite{levien2021non}, and more.

\begin{acknowledgments}
We thank Hanna Salman and Fangwei Si for valuable discussions. This work was supported by National Science Foundation Grant Nos.\ PHY-2118561 and DMS-2245816.
\end{acknowledgments}


\onecolumngrid
\newpage

\beginsupplement

\begin{center}
    {\bf SUPPLEMENTAL MATERIAL}
\end{center}
\vspace{.4in}

\begin{table}[h]
\begin{center}
{\bf Variables} $\qquad\qquad\qquad\qquad$
{\bf Parameters} $\qquad\qquad\qquad\qquad\qquad\qquad$
{\bf Defined quantities} $\qquad\qquad$\\
\begin{tabular}{|c | l |} 
 \hline
 $t$ & Time \\
 \hline
 $n$ & Generation \\
 \hline
 $b_n$ & Cell birth size \\
 \hline
 $t_n$ & Cell division time \\
 \hline
 $s_n(t)$ & Cell size \\
 \hline
 $x_n(t)$ & Molecule number \\
 \hline
\end{tabular}
\begin{tabular}{|c | l |} 
 \hline
 $\alpha$ & Cell growth rate \\ 
 \hline
 $\beta$ & Homeostasis parameter \\ 
 \hline
 $\nu$ & Size-independent production rate \\ 
 \hline
 $\mu$ & Size-dependent production rate \\ 
 \hline
 $\lambda$ & Degradation rate \\ 
 \hline
 $x_*$ & Molecule number threshold \\ 
 \hline
\end{tabular}
\begin{tabular}{| l | l |} 
 \hline
 $\epsilon_n\equiv\ln(b_n/\bar{b})$ & Scaled, zero-mean birth size\\ 
 \hline
 $\delta_n\equiv\alpha t_n-\ln2$ & Scaled, zero-mean division time\\ 
 \hline
 $\gamma\equiv\nu/\mu\bar{b}$ & Production rate ratio\\ 
 \hline
 $\rho\equiv\lambda/\alpha$ & Degradation per growth\\ 
 \hline
 $k\equiv\nu+\mu\bar{b}$ & Total production rate \\ 
 \hline
 $r\equiv2^\rho$, $g\equiv\gamma \rho + \gamma$ & For notational convenience \\ 
 \hline
\end{tabular}
\caption{Symbols used in this study.}
\label{parmatersdef}
\end{center}
\end{table}

\section{Master Equation and Generating Function}
Eqs.\ \ref{s} and\ \ref{xdot} define the average dynamics of our model. With size growing exponentially with rate $\alpha$ and protein number $x$ growing scholastically with birth and death rates, $\nu+\mu s$ and $\lambda x$ respectively. Our goal is to find the effect of intrinsic molecular noise on size statistics. We start by writing the birth-death master equation for $x$ 
\begin{equation}
\label{master}
\partial_t p(x,t)=t^+(x-1) p(x-1,t) + t^-(x+1) p(x+1,t) - (t^+(x)+t^-(x))p(x,t),
\end{equation}
where $t^+(x)$ and $t^-(x)$ are the transition rates for the processes $x\rightarrow x+1$ and $x\rightarrow x-1$ respectively. The transition rates can be read off from Eq.\ \ref{xdot} to be
\begin{equation}
\label{rates}
\begin{aligned}
t^+(x)&=\nu+\mu s,\\
t^-(x)&=\lambda x,
\end{aligned}
\end{equation}
and Eq.\ \ref{master} becomes
\begin{equation}
\partial_t p(x,t)=(\nu+\mu s) p(x-1,t) + \lambda (x+1) p(x+1,t) - (\nu+\mu s+\lambda x)p(x,t).
\end{equation}
After inserting the solution for Eq.\ \ref{s}, we get
\begin{equation}
\label{master2}
\partial_t p(x,t)=(\nu+\mu b_ne^{\alpha t}) p(x-1,t) + \lambda (x+1) p(x+1,t) - (\nu+\mu b_ne^{\alpha t}+\lambda x)p(x,t).
\end{equation}
We define a generating function $G(s,t)$ as
\begin{equation}
\label{generating}
G(s,t)=\sum_{x=0} s^x p(x,t),
\end{equation}
then, differentiate w.r.t time
\begin{equation}
\label{generating2}
\partial_t G(s,t)=\sum_{x=0} s^x \partial_t p(x,t).
\end{equation}
Substituting Eq.\ \ref{master} in\ \ref{generating2}, we get
\begin{equation}
\label{generating3}
\partial_t G(s,t)=\sum_{x=0} s^x [(\nu+\mu b_ne^{\alpha t}) p(x-1,t) + \lambda (x+1) p(x+1,t) - (\nu+\mu b_ne^{\alpha t}+\lambda x)p(x,t)].
\end{equation}
Each term in the RHS of Eq.\ \ref{generating3} can be rewritten as
\begin{align}
&\sum_{x=0} s^x p(x-1,t)=\sum_{x=1} s^x p(x-1,t)=\sum_{x=0} s^{x+1} p(x,t)=s G(s,t),\\
&\sum_{x=0} s^x (x+1)p(x+1,t)=\sum_{x=1}s^{x-1}xp(x,t)=\partial_s \sum_{x=1} s^x p(x,t)=\partial_s \Big[\sum_{x=0} s^x p(x,t) - p(0,t)\Big]=\partial_s G(s,t),\\
&\sum_{x=0} s^x xp(x,t)=s \partial_s \sum_{x=0}s^x p(x,t)=s \partial_s G(s,t),
\end{align}
where the boundary condition $p(x<0,t)=0$ is assumed. Therefore, Eq.\ \ref{generating3} becomes
\begin{equation}
\label{generating4}
\begin{aligned}
\partial_t G(s,t)&=(\nu+\mu b_ne^{\alpha t}) s G(s,t)+ \lambda \partial_s G(s,t) - (\nu+\mu b_ne^{\alpha t})G(s,t)-\lambda s \partial_s G(s,t)\\
&=(\nu+\mu b_ne^{\alpha t}) (s-1) G(s,t)-\lambda(s-1) \partial_s G(s,t).
\end{aligned}
\end{equation}
The solutino to this PDE is
\begin{equation}
\label{generatingsol}
G(s,t)=\exp \Big(\frac{s \nu}{\lambda} + \frac{e^{\alpha t} b_n \mu (s-1)}{\lambda+\alpha} \Big) F(t+\frac{1}{\lambda} \ln{\frac{1}{s-1}}),
\end{equation}
where $F$ is an unknown function. Since at $t=0$, molecules number $x(0)=x_0$, the distribution of $x$ at $t=0$ is $p(x,0)=\delta_{x x_0}$. Thus, the condition
\begin{equation}
\label{boundary}
G(s,0)=\sum_{x=0} s^x \delta_{x x_0}=s^{x_0},
\end{equation}
which enables us to find $F$,
\begin{equation}
F(\frac{1}{\lambda} \ln{\frac{1}{s-1}})=s^{x_0} \exp \Big(\frac{-s \nu}{\lambda} + \frac{ b_n \mu (1-s)}{\lambda+\alpha} \Big).
\end{equation}
we define the parameter $s^\prime=\frac{1}{\lambda} \ln{\frac{1}{s-1}}$ and solve for $s$ in terms of $s^\prime$ to find $F(s^\prime)$, resulting in
\begin{equation}
F(s^\prime)=(1+e^{-s^\prime \lambda})^{x_0} \exp\Big( \frac{-\nu}{\lambda} (1+e^{-s^\prime \lambda})  - \frac{b_n \mu e^{-s^\prime \lambda}}{\lambda+\alpha}  \Big).
\end{equation}
Hence,
\begin{equation}
F(t+\frac{1}{\lambda} \ln{\frac{1}{s-1}})=(1+e^{- \lambda t} (s-1))^{x_0} \exp\Big( \frac{-\nu}{\lambda} (1+e^{- \lambda t} (s-1))  - \frac{b_n \mu e^{- \lambda t} (s-1)}{\lambda+\alpha}  \Big),
\end{equation}
which when plugged in\ \ref{generatingsol} gives us the full solution for the generating function
\begin{equation}
\label{generatingfunc}
G(s,t)=\exp \Big( \frac{b_n \mu (s-1)}{\lambda+\alpha} (e^{\alpha t} - e^{-\lambda t}) + \frac{\nu}{\lambda} (s-1-e^{-\lambda t} (s-1)) \Big) (1+e^{- \lambda t} (s-1))^{x_0}.
\end{equation}

\section{Molecular Noise}

Now that we derived the generating function, we can use it to derive the moments for $x$. The first and second moments can be derived using the generating function as follows
\begin{align}
&\partial_s G(s,t)=\sum_{x=0}x s^{x-1}p(x,t),\\
&\partial^2_s G(s,t)=\sum_{x=0}x(x-1) s^{x-2}p(x,t),
\end{align}
for $s=1$ this becomes
\begin{align}
&\partial_s G(s,t)|_{s=1}=\sum_{x=0}x p(x,t)=\langle x \rangle,\\
&\partial^2_s G(s,t)|_{s=1}=\sum_{x=0}x(x-1)p(x,t)=\sum_{x=0}x^2 p(x,t) - \sum_{x=0}xp(x,t)= \langle x^2 \rangle - \langle x \rangle.
\end{align}
Thus, the variance given by
\begin{equation}
\sigma_x^2=\langle x^2 \rangle - \langle x \rangle^2=\partial^2_s G(s,t)|_{s=1}+\partial_s G(s,t)|_{s=1}-(\partial_s G(s,t)|_{s=1})^2,
\end{equation}
which results in
\begin{equation}
\label{xnoise1}
\sigma_{x|b_n}^2=\frac{e^{-2\lambda t}}{\lambda (\lambda + \alpha)} \Bigg[x_0 (\lambda^2 + \alpha \lambda) (e^{\lambda t}-1) +(\alpha \nu + \lambda \nu) e^{\lambda t} (e^{\lambda t} -1)+\lambda b_n \mu e^{\lambda t} (e^{(\lambda + \alpha) t}-1) \Bigg].
\end{equation}
We use rescaled parameters defined in table \ref{parmatersdef}, Eq.\ \ref{xnoise1} becomes
\begin{equation}
\label{xnoise2}
\sigma_{x|b_n}^2=\frac{e^{-2\rho \alpha t}}{\rho (\rho + 1)} \Bigg[x_0 (\rho+\rho^2) (e^{\rho \alpha t}-1) +\frac{k}{\alpha}\frac{\gamma}{1+\gamma}(1+\rho) e^{\rho \alpha t} (e^{\rho \alpha t} -1)+\rho b_n \frac{k}{\alpha}\frac{1}{(1+ \gamma)} e^{\rho \alpha t} (e^{(\rho + 1) \alpha t}-1) \Bigg].
\end{equation}

\section{Homeostasis Parameter $\beta$}
Next, we consider times $t$ near division and write the expression for $\bar{x}_n(t)$ in the main text in terms of, $\delta=\alpha t-\ln{2}$ and $\epsilon_n = \ln{(b_n/\bar{b})}$, the deviations from the average division phase $\alpha t$ and the average birth size respectively, and expand to first order in $\epsilon_n$ and $\delta$. We find
\begin{align}
\label{xdev}
\bar{x}_n(\delta) =&\ x_0 2^{-\rho}(1-\rho \delta) + \frac{k \gamma}{\alpha \rho (1+\gamma)} [1- 2^{-\rho}(1-\rho \delta)] + \frac{k(1+\epsilon_n)}{\alpha(1+\gamma)(1+\rho)} [2(1+\delta)-2^{-\rho}(1-\rho \delta)]
\nonumber\\ \approx&\ c_0+c_1\epsilon_n+c_2\delta,
\end{align}
where $x_0=x_{*}/2$, $x_{*}$ is the abundance threshold, $c_0=x_0 2^{-\rho}+k\gamma[\alpha\rho(1+\gamma)]^{-1} [1 -2^{-\rho}]+k[\alpha(1+\gamma)(1+\rho)]^{-1}[2-2^{-\rho}]$, $c_1=k[\alpha(1+\gamma)(1+\rho)]^{-1}[2-2^{-\rho}]$ and $c_2=k\gamma 2^{-\rho}[\alpha(1+\gamma)]^{-1} + k[\alpha(1+\gamma)(1+\rho)]^{-1}[2+\rho 2^{-\rho}]-x_0\rho2^{-\rho}$. At division, we define $\delta=\delta_n$ and take the average of Eq.\ \ref{xdev}, we find $c_0=x_{*}$, and Eq.\ \ref{xdev} becomes,
\begin{align}
\label{xdev2}
\bar{x}_n(\delta) \approx x_{*}+c_1\epsilon_n+c_2\delta.
\end{align}
Consequently, Eq.\ \ref{xdev2} at division reads
\begin{align}
\label{xdev3}
0 \approx c_1\epsilon_n+c_2\delta_n,
\end{align}
which, when compared to Eq.\ \ref{delta} without the noise, reads 
\begin{align}
\label{betamodel}
\beta=\frac{c_1}{c_2}=\frac{(2 r -1)^2}{4 r^2 + (g-2) r},
\end{align}
where $g=\gamma \rho + \gamma$ and $r=2^\rho$. Now we have a definition of $\beta$ in terms of our model parameters which allows us to understand the mapping of our model dynamics to different homeostasis regimes. 
\section{Size Noise}
Noise in the growth factor at division, $\delta_n$, ``timing noise" is directly contributing to size noise from Eq.\ \ref{size_noise}. The effect of molecule number noise on size noise can be seen directly from its relation to timing noise given by Eq.\ \ref{timing_noise}. To derive Eq.\ \ref{timing_noise}, we invoke the approximation $\sigma^2_{\delta_n|\epsilon_n} \approx \big(\left.\frac{\partial \bar{x}_n}{\partial\delta_n}\right|_{\bar{\delta_n}=0}\big)^{-2}\sigma^2_{x_n|\epsilon_n}$, which assumes that higher order terms of $\delta_n$ are negligible, or equivalently growth factors at division are tightly distributed around the average. We should also indicate that the timing noise for the population is the average of the noise in the growth factor over size. This results in
\begin{equation}
\label{timing_noise_s}
\sigma^2_{\eta} = \left\langle\sigma^2_{\delta_n|\epsilon_n}\right\rangle
\approx \left\langle\left(\left.\frac{\partial \bar{x}_n}{\partial\delta}\right|_{\delta=0}\right)^{-2}\sigma^2_{x_n|\epsilon_n}\right\rangle.
\end{equation}
From Eq.\ \ref{xdev2} we find $\big(\left.\frac{\partial \bar{x}_n}{\partial\delta}\right|_{\delta=0}\big)^{-2} =1/c_2^2$, and Eq.\ \ref{timing_noise_s} becomes
\begin{equation}
\label{timing_noise_s2}
\sigma^2_{\eta} \approx \frac{\left\langle \sigma^2_{x_n|\epsilon_n}\right\rangle}{c_2^2}.
\end{equation}
The noise $\sigma^2_{x_n|\epsilon_n}$ can be found by writing down Eq.\ \ref{xnoise2} in terms of $\epsilon_n$ and $\delta_n$, which we find to be
\begin{align}
\label{xnoise3}
\sigma_{x|\epsilon_n}^2= c_1 \epsilon_n + (c_2 + 2 \rho x_0 2^{-2 \rho}) \delta_n + x_{*} - x_0 2^{-2 \rho},
\end{align}
given that $\left\langle \epsilon_n \right\rangle=0$ and $\left\langle \delta_n \right\rangle=0$, Eq.\ \ref{timing_noise_s2} becomes
\begin{equation}
\label{timing_noise_s3}
\sigma^2_{\eta} \approx \frac{x_{*} - x_0 2^{-2 \rho}}{c_2^2}.
\end{equation}
Substituting Eqs.\ \ref{timing_noise_s3} and\ \ref{beta} in Eq.\ \ref{size_noise}, size noise becomes
\begin{equation}
\label{size_noise_s1}
\frac{\sigma^2_b}{\bar{b}^2} \approx \sigma^2_\epsilon = \frac{x_{*} - x_0 2^{-2 \rho}}{c_1(2c_2-c_1)}.
\end{equation}

We found numerically that the curve of minimum size noise is parametrized by $\rho\to0$ when $0<\beta\le1/2$ and $\gamma\to0$ when $1/2<\beta<1$. To find the theoretical bound on size noise, we solve Eq.\ \ref{betamodel} in each branch for the other non-zero parameter in terms of $\beta$. This gives us
\begin{align}
\label{limits_1}
&\rho=\to0,\ \gamma\to\frac{1}{\beta}-2,\ \beta\le1/2 \\
&\rho\to\frac{\ln{(1/(1-\beta))}}{\ln{2}}-1,\ \gamma\to0,\ \beta>1/2 
\label{limits_2}
\end{align}
which when combined with Eq.\ \ref{size_noise_s1} results in the final equation for the theoretical noise bound,
\begin{equation}
\label{min_s}
\frac{\sigma^2_b}{\bar{b}^2} \ge \frac{\alpha/k}{\beta(2-\beta)}
\begin{cases}
(1-\beta)[\beta+(1-2\beta)\ln2] & \beta\le1/2 \\
c(2\beta^2-4\beta+1)\ln(1-\beta) & \beta>1/2.
\end{cases}
\end{equation}
\section{Simulation and Data Analysis}
We simulated the stochastic dynamics of $x$ using the stochastic simulation algorithm, then calculated the size noise and homeostasis parameter $\beta$. Each point in the simulation shown in Fig.\ \ref{mean}(e) corresponds to a different value of $\gamma$, $\rho$, and $\alpha/k$, sampled uniformly in log space. We expect our simulation to obey the theoretical bound with some deviation instances due to the approximations we made in Eqs.\ \ref{expand} and \ref{timing_noise}.

We performed analysis on publicly available from six microfluidic studies on {\it E.\ coli} referenced in the main text. All data sets typically contained size, time and growth rate measurements. In references \cite{vashistha2021non,wang2010robust}, where the available data did not contain rate measurements, we calculated $\alpha$ for each generation by assuming exponential growth of cells [Eq.\ \ref{s}]. Solving for $\alpha$, we find the relation
\begin{equation}
\label{rate_exp}
\alpha=\frac{1}{T}\ln{\frac{s_n(T)}{b_n}},
\end{equation}
where $T$ is generational time, $s_n(T)$ is division size, and $b_n$ is birth size. To obtain the value of $\beta$ from experiments we performed linear regression analysis using $\epsilon_n$ and $\delta_n$ calculated from data. According to Eq.\ \ref{delta}, $\beta$ is the negative value of the resulting slope.

Error bars in all figures represent standard error (SE) in the data. For $\alpha$, SE was obtained directly from the data by calculating $\sigma_{\alpha}/\sqrt{n}$. In the case of $\beta$, SE was calculated for the fit using MATLAB. In order to calculate SE for the $CV^2$ ($\sigma_b^2/\bar{b}^2$) in Fig.\ \ref{noise}(b), we performed bootstrapping on the data. For each data set, we randomly sampled $N$ data points that we used to calculate $\sigma_b^2/\bar{b}^2$ starting from $N=5$. We then repeated the random sampling $50$ times for the same $N$ and calculated SE for the obtained $50$ values of $\sigma_b^2/\bar{\alpha}\bar{b}^2$. After that, we increased the value of $N$ gradually and repeated the procedure, eventually we could interpolate the SE value for the total number of available data points.

\section{Extrinsic noise sources}
\begin{figure}
\includegraphics[width=0.8\columnwidth]{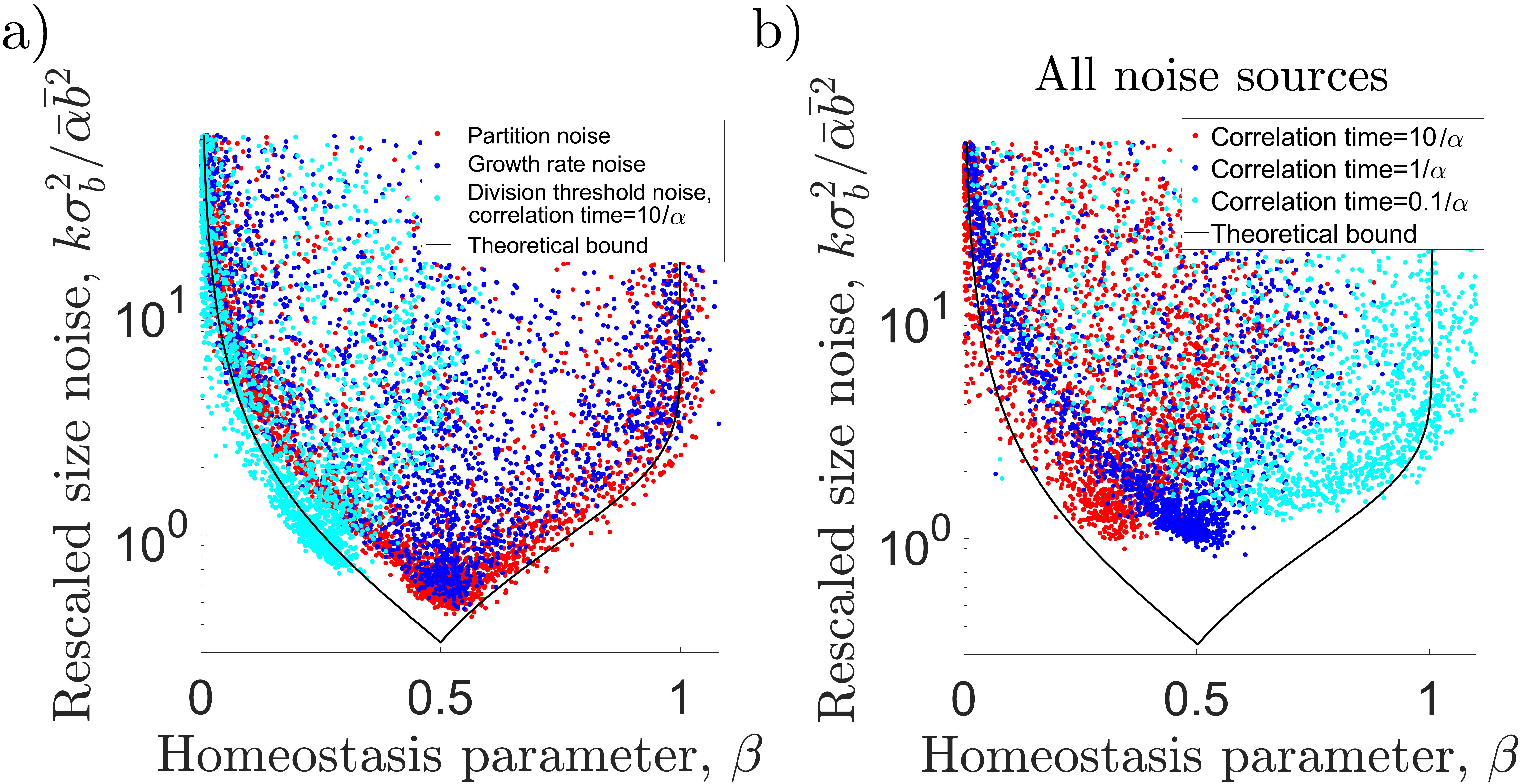}
\caption{(a) Rescaled size noise ($CV^2$) vs.\ homeostasis parameter $\beta$ from simulations showing the effect of individual extrinsic source of noise. (b) The effect of all sources of noise combined on size noise.}
\label{noise_robust}
\end{figure}

Other sources of noise are expected to affect the distribution and contribute to noise. We tested the robustness of our results to three noise sources: (1) growth rate noise, (2) partitioning noise of the size control molecules upon division, and (3) noise in the copy-number threshold that initiates division. We again used the stochastic simulation algorithm to simulate the stochastic dynamics of accumulation of the size control molecule. Results are shown in Fig.\ \ref{noise_robust}.

Noise in the growth rate was sampled each generation and simulated as a gaussian white noise with noise strength of 20\% of the average , with noise strength chosen consistently with the experimental data (see Fig. \ref{noise} a). The partitioning noise was simulated by distributing molecules upon division according to a binomial distribution between the two daughter cells. Copy number threshold noise was accounted for as intrinsic fluctuations with a given correlation time, which we varied, and simulated using the stochastic simulation algorithm.

We find that including all noise sources does not change the main results of our model, namely, the existence of a minimum size noise boundary, and a globally minimum size noise that is away from the sizer in contrast to the standard model (see Fig.\ \ref{noise_robust}). We found that, depending on the correlation time of the fluctuations in the division threshold, data will shift more towards the timer (high correlation time) or the sizer (small correlation time). This observation is consistent with the results reported before in \cite{luo2023stochastic}. In all cases, our predicted bound was obeyed, with the simulation data tending to increase upon adding more noise sources. In all cases, a clear minimum robustly existed for the size noise, away from the sizer.

\section{Fixing $x_*$ instead of $k$}
\begin{figure}
\includegraphics[width=0.5\columnwidth]{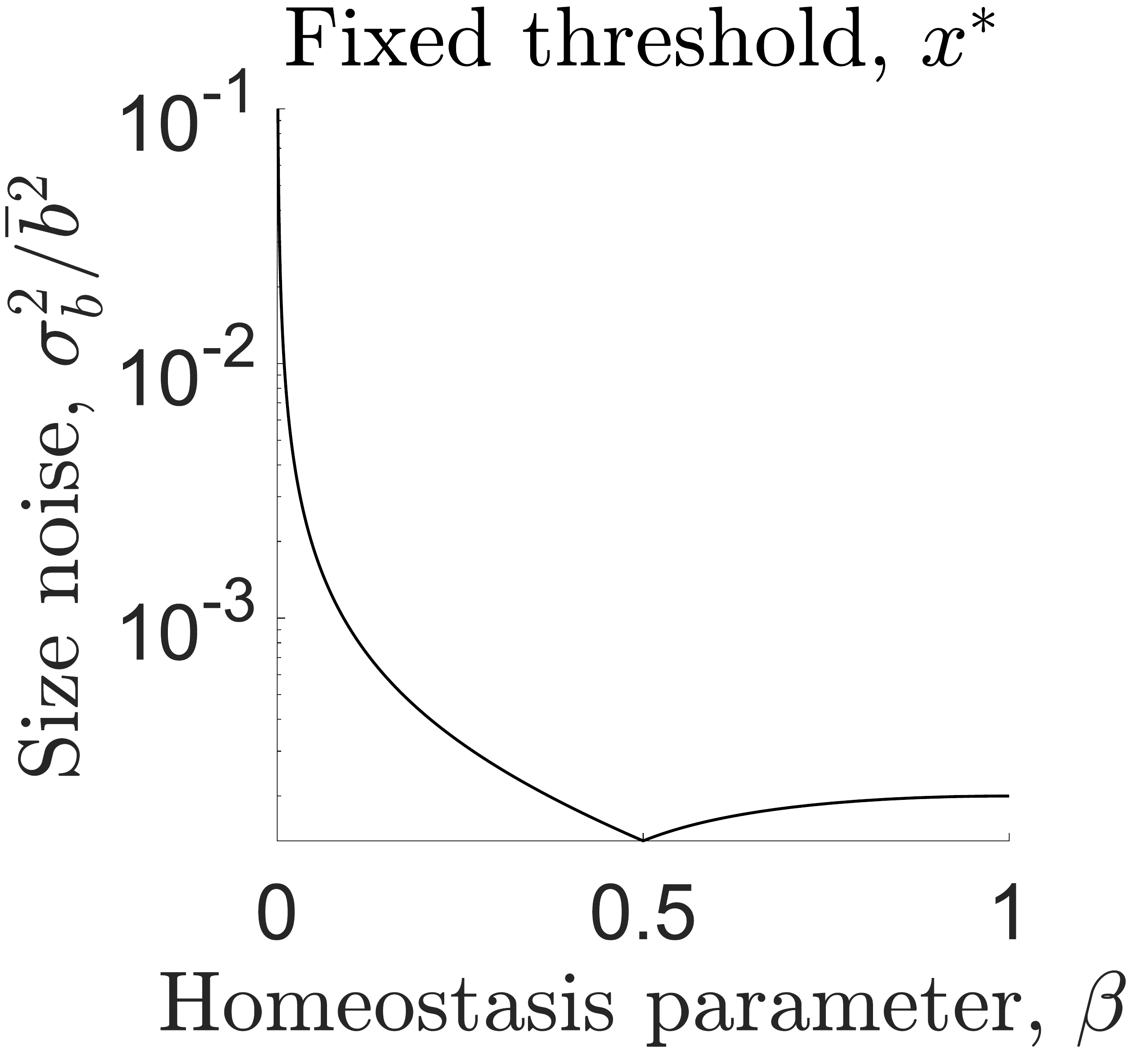}
\caption{Minimum size noise vs $\beta$ for a fixed threshold $x^*=5000$. The global minimum is still at the adder ($\beta=1/2$).}
\label{min_xstar_fig}
\end{figure}

For quantity-based mechanisms, where a component must reach a threshold, the sizer limit is achieved by making the degradation of that component much faster than the growth of the cell, so that the component tracks the cell volume at all times. To compensate for the fast degradation, either the production of the component must be high, or the threshold must be low. In our model, we assume a fixed cost (constant $k$), therefore, $x^{*}\to0$ as $\beta\to1$. If we relax this assumption and instead require that $x^{*}$ is the fixed quantity, production $k$ will change with $\beta$ to meet the threshold, signifying different resource allocation. Using Eqs.\ \ref{xstar}, \ref{limits_1}, and \ref{limits_2}, we find $k$, for a fixed $x^{*}$, to be
\begin{equation}
\label{k}
k= \frac{ x^{*}\alpha}{2}
\begin{cases}
(1-\beta)/(\beta+\ln2-2\beta\ln2) & \beta\le1/2 \\
\ln[(1-\beta)^{-1}]/\ln2 & \beta>1/2.
\end{cases}
\end{equation}
We can derive the size noise under the fixed $x^{*}$ assumption by substituting Eq.\ \ref{k} in Eq.\ \ref{min}, we find
\begin{equation}
\label{min_xstar}
\frac{\sigma^2_b}{\bar{b}^2} \ge \frac{1}{x^{*}\beta(2-\beta)}
\begin{cases}
2(\beta+\ln2(1-2\beta))^{2} & \beta\le1/2 \\
(4\beta-2\beta^2-1) & \beta>1/2.
\end{cases}
\end{equation}
Eq.\ \ref{min_xstar} is plotted in Fig.\ \ref{min_xstar_fig}, and it shows that while the size noise no longer diverges at the sizer limit ($\beta=1$), it still increases monotonically in going from the adder ($\beta=1/2$) to the sizer. The reason is that, while the standard model would predict a decrease in the size noise according to the homeostasis factor $H(\beta)=\beta^{-1} (2-\beta)^{-1}$ (Fig.\ \ref{mean}b), this decrease is outweighed by an increase in the molecule number noise from $x^*/2$ to $x^*$ (described after Eq.\ \ref{timing_noise}). Specifically, from adder to sizer, the former effect decreases size noise by $H(1)/H(1/2) =3/4$, while the latter effect increases size noise by $x^*/(x^*/2)=2$.

Eq.\ \ref{k} also shows that a fixed threshold implies a large production rate in the sizer limit. We see that for a pure sizer ($\beta=1$) $k$ is infinite, but even near the sizer it can be large. For example, taking a conservative value of the growth rate from Fig \ref{noise}a, 
$\alpha\approx0.5/hr$, and taking the estimate $x^{*}\approx5000$ for the candidate accumulator protein FtsZ (\cite{feucht2001cytological}, Table 1), we obtain $k\approx20$/min for the adder ($\beta=1/2$). This value already matches the typical rates of transcription initiation and translation initiation in E. coli, both around $20$/min (\cite{kennell1977transcription}, Table 2). As we approach the sizer, the production rate increases: for $\beta=0.9$ we have $k\approx70$/min, and for $\beta=0.99$ we have $k\approx140$/min. Thus, we can see that under the fixed-threshold hypothesis, a sizer strategy could easily require a protein production rate that exceeds the limits set by the basic constraints of transcription and translation.

In summary, by fixing the protein abundance threshold rather than the protein production rate, we find that: 1) a qualitative minimum at $\beta=1/2$ still holds, and 2) the protein production rate required to achieve sizer control ($\beta\to1$) can readily exceed the fundamental speed limit of transcription and translation. Quantitatively, Fig.\ \ref{min_xstar_fig} shows that the minimum at $\beta=1/2$ is shallow at larger $\beta$. Thus, on the basis of (1) alone, it could be concluded that $0.5 < \beta < 1$ is largely equivalent from a noise perspective, and that requirements other than noise minimization may set the value of $\beta$ within this range. Nonetheless, it remains true that (2) is a fundamental observation about protein production limits and is sufficient to suggest that a pure sizer is unlikely to be achieved under a constant threshold protocol. In general, we conclude that our main finding, that size noise is not minimized by the sizer strategy, is robust to the precise optimization protocol.

\end{document}